\documentclass[showpacs,aps,pra,floatfix,reprint,superscriptaddress,footinbib]{revtex4-1}
\usepackage{xfrac} 
\usepackage{amssymb,amsfonts,amsmath}
\usepackage{braket}
\usepackage{graphicx}
\usepackage{verbatim}
\usepackage{etoolbox}
\usepackage[utf8]{inputenc} 
\usepackage{mathtools}
\usepackage{soul,xcolor,colortbl}
\usepackage{setspace}
\usepackage{multirow}
\usepackage{makecell}
\usepackage[normalem]{ulem}
\usepackage{scrextend}

\usepackage{hyperref} \hypersetup{colorlinks=true,citecolor=blue,linkcolor=blue,urlcolor=blue} 

\definecolor{pranab_green}{rgb}{0.31,0.53,0.10}
\definecolor{pranab_red}{rgb}{0.85,0.23,0.11}

\usepackage{bm} 
\usepackage{array}
\newcolumntype{C}[1]{>{\centering\arraybackslash}p{#1}}\usepackage{soul}
\definecolor{Gray}{gray}{0.85}

\usepackage{color} 
\usepackage{morefloats}

\definecolor{Gray}{gray}{0.9}
\definecolor{LightCyan}{rgb}{0.88,1,1}
\def\AFLOW{{\small AFLOW}}
 
\def\AFLOWCHULL{{\small AFLOW-CHULL}} 
\def\RESTAPI{{\small REST-API}}    
\def\AFLUX{{\small AFLUX}}

\def\JSON{{\small JSON}}

\def\DHNN{{\Delta H \left[N|\left\{1,\cdots\!,N\!\!-\!\!1\right\}\right]}}

\makeatletter \renewcommand\frontmatter@abstractwidth{\dimexpr\textwidth\relax} \makeatother 

\begin{document}
\title{Unavoidable disorder and entropy in multi-component systems}

\author{Cormac Toher}
\affiliation{Dept. Mechanical Engineering and Materials Science, Duke University, Durham NC, 27708, USA}
\author{Corey Oses}
\affiliation{Dept. Mechanical Engineering and Materials Science, Duke University, Durham NC, 27708, USA}
\author{David Hicks}
\affiliation{Dept. Mechanical Engineering and Materials Science, Duke University, Durham NC, 27708, USA}
\author{Stefano Curtarolo}
\email[]{stefano@duke.edu}
\affiliation{Dept. Mechanical Engineering and Materials Science, Duke University, Durham NC, 27708, USA}

\date{\today}

\begin{abstract}
  \noindent
  The need for improved functionalities is driving the search for more complicated multi-component materials.
  {Despite the factorially increasing composition space, ordered compounds with 4 or more species are rare.}
  {Here, we unveil} the competition between {the gain in} enthalpy and entropy {with increasing number of species} by statistical analysis of {the \AFLOW} data repositories.
  A threshold in the number of species is found {where entropy gain exceeds enthalpy gain}.
  Beyond that, enthalpy can be neglected, and disorder --- complete or partial --- is unavoidable.
\end{abstract}
\maketitle

\noindent
The formation of materials in the tangible world, enjoying easy-life at room temperature and pressure, is governed by the Gibbs free energy.
Enthalpy usually makes up a {large} part of it, and, for the last two decades, the search for new materials has {mostly focused} on enthalpy optimization.
A plethora of methods were proposed: e.g. cluster expansion \cite{deFontaine_ssp_1994}, genetic- \cite{Bush_JMC_GA_1995} and combinatorial algorithms \cite{nmatHT}.
In trying to cure the Maddox curse \nocite{Maddox}\footnote{Maddox \cite{Maddox}: ``One of the continuing scandals in the physical sciences is that it remains in 
  general impossible to predict the structure of even the simplest crystalline solids from a knowledge of their chemical composition''.},
these approaches were all somewhat successful in predicting and discovering systems made up of few species.
At least this was the impression: polymorphs were still able to escape from the enthalpic jail!
While the need for functionalities was driving the search for more complicated materials, something unexpected happened:
the discovery of high-entropy alloys \cite{Gao2015highdesign, Widom_Multicomponent_JSP_2017}.
The ideal scenario of phase transitions --- in which precursors constantly seek the most stable configuration --- was shattered by the acceptance that disorder is useful.
Skeptics {considered} high-entropy alloys only a remake of solid-solutions. In the meantime, new families appeared: entropy-stabilized oxides, high-entropy carbides and borides. 
Each carrying a panoply of unexpected properties. 

\begin{figure*}
  \centerline{\includegraphics[width=0.98\textwidth]{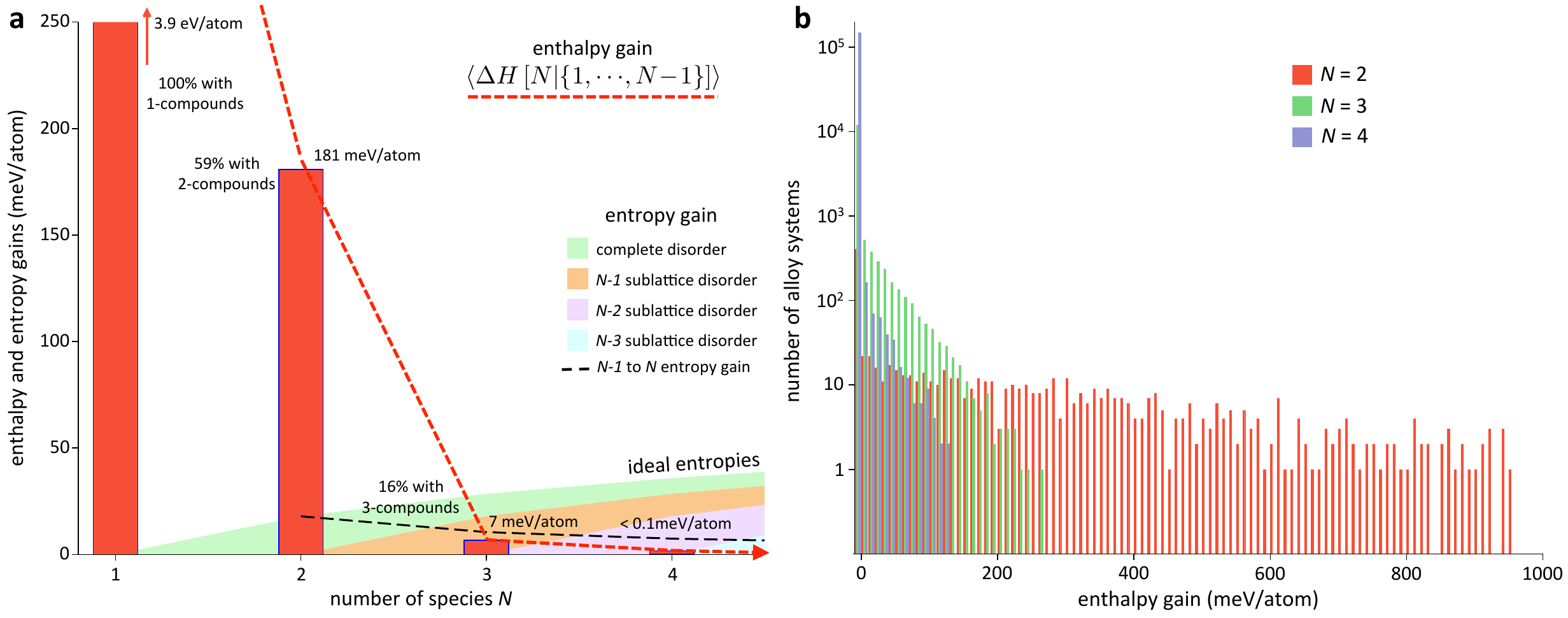}} 
  \vspace{-2mm}
  \caption{\small
    {\bf Enthalpy and entropy gains.}
    (\textbf{a}) Expectation of enthalpy-gain $\left<\DHNN\right>$ and ideal entropy contributions as functions of number of species, $N$.
    Even at room temperature, entropy eventually overwhelms enthalpy in controlling phase stability of multi-component systems.
    {(\textbf{b}) Enthalpy gain distribution for $N=2,3,4$; gains decrease with increasing number of species.}} 
  \label{fig1}
\end{figure*}

\noindent
{\bf Are entropic materials the exception?}
The question requires the disentanglement of the ``{\it enthalpy versus entropy}'' dichotomy.
Big-data statistical analysis gives the answer.
Let us consider the {\small AFLOW}.org {\it ab-initio} materials repository \cite{aflowMRS} and analyze its compounds' energies with the associated tools \cite{aflux,aflowCHULL}. 
The recursive formation enthalpy ``gain'' of an $N$-species ordered compound (called $N$-compound) with respect to combinations of $\{1,\cdots\!,N\!-\!1\}$-species ordered sub-components,
$\DHNN$, is defined as the energetic distance of the enthalpy of the $N$-compound, $H\left[N\right]$, below the $\{1,\cdots\!,N\!-\!1\}$ convex-hull hyper-surface
$H_{\rm hull} \left[\{1,\cdots\!,N\!\!-\!\!1\}\right]$ generated from its $\{1,\cdots\!,N\!\!-\!\!1\}$-species components: $\DHNN \equiv  H_{\rm hull} \left[\{1,\cdots\!,N\!\!-\!\!1\}\right]-H\left[N\right]$ if $H\left[N\right]<H_{\rm hull} \left[\{1,\cdots\!,N\!\!-\!\!1\}\right]$ 
\footnote{\label{binaries} For binary compounds, $\Delta H\left[2|1\right]$ is equivalent to the usual formation enthalpy. 
  For $N>2$, the formation enthalpies can be written as sums of the recursive gains.}
and zero otherwise 
\footnote{$\DHNN$ is conceptually similar to the stability criterion $\delta _{\rm sc}$ described in Ref. \protect\onlinecite{aflowCHULL}, except with all $N$-species entries removed from the pseudo-hull.}.
Physically, $\DHNN$ represents the enthalpy gain to create an ordered $N$-compound out of all the combinations of ordered $\left\{ 1-, \cdots\!, N\!-\!1\right\}$-ones.
Gain expectations can be obtained by the analysis of {metal alloy phase diagrams ``constructible''} out of the {\small AFLOW}.org data.
{$\left<\Delta H\right>$ is shown in Figure \ref{fig1}(a) for $N=2,3,4$, and as cohesive energy for $N=1$.}


\noindent
{\bf Finding new $N$-compounds becomes difficult. }
While {59\% (588)} of the binaries show enthalpy gain by forming {a total of 1995} 2-compounds, only { 16\% (2237)} of ternaries produce
{3040} 3-compounds {and just 426 (0.3\%) quaternaries form 4-compounds}.
Phase separation is overwhelming.
Contrary to common belief, the trend indicates that despite the combinatorially-growing number of potential configurations, 
finding accommodating prototypes becomes harder with increasing $N$. 

\noindent
{\bf Entropy takes over.}
The surviving combinations of species forming new compounds become vulnerable with increasing $N$.
The expectation of $\Delta H$ is $\sim$181~meV/atom for binaries.
It drastically reduces to $\sim$7~meV/atom for ternaries, practically disappearing ($< 0.1$~meV/atom) for quaternaries.
{434 binary, 192 ternary but only 10 quaternary systems have a large enthalpy gain (defined as exceeding 100~meV/atom):
the frequency distributions of enthalpy gains for all systems are displayed in Figure \ref{fig1}(b).}
The numbers are compared to disorder promotion where the ideal scenario is considered for simplicity \cite{McQuarrie}.
Figure \ref{fig1}(a) shows the ideal entropic contribution at room temperature for the fully ($N$) and partially disordered ($N-1$, $N-2$, $N-3$) systems. 
{Since the $N-1$ component systems could also feature disorder, \textit{i.e.} the $N$ component system could phase-separate into multiple disordered phases, the $\log(N) - \log(N-1)$ entropy gain is also plotted (black dashed line).
Note that the latter already exceeds the enthalpy gain for $N=3$.}
The reducing probability of having $N$-compounds --- also recursively applying to sub-components ---
compounded with the drastically decreasing enthalpic gain and the monotonic entropic promotion, indicates that around the threshold $N\sim4$, entropy takes over the stability of the system. 

\noindent
{\bf Beyond metals.}
The analysis focuses on the extensive \AFLOW\ set of metallic compounds{; a similar analysis for non-metals requires formation enthalpy corrections \cite{Friedrich_CCE_2019}}.
Mixing different types of bonding can change the entropy threshold by adding additional enthalpic stabilization, leading to a shift of the onset of the entropic promotion ($N$ increases).
It is not a coincidence: entropy stabilized oxides were discovered with 5 metallic species mixed with oxygen \cite{curtarolo:art99}, and high-entropy-carbides {and borides} were
also synthesized with 5 metallic species plus carbon \cite{curtarolo:art140} {or boron \cite{Gild_borides_SciRep_2016}}. There, oxygen{, carbon and boron} are ``spectator'' species.
The devil is in the details: reciprocal systems might be able to reduce 
the number of the mixing species, thanks to the additional entropic stabilization of every equivalent sublattice.

\noindent
{\bf The future.}
Ramifications of the analysis are intriguing.
Disordered systems can lead to {remarkable} properties \cite{Gludovatz_hea_mech_properties, HEAprop2, Braun_ESO_AdvMat_2018}
--- unexpected from the homologous crystalline counterparts --- 
enabling revolutionary technologies.
We just need to search for materials with a different mindset.
{\bf I.} 
The search for multi-component systems, performed with enthalpy optimization, is futile. 
With increasing $N$, the discovery probability decreases, and, even if an ordered material is found, it would be overwhelmed by engulfing disorder.
High-entropy promotion occurs at $\sim$4 mixing species ($\sim$5+ with different bond types). 
{\bf II.} 
{Sluggish kinetics at low temperatures can be a blessing \cite{BaluffiAllenCarter_book}. 
Disorder is hard to cure, and high entropy solid-solutions synthesized at high-temperature can survive low-temperature phase separation
\cite{curtarolo:art140, BaluffiAllenCarter_book} while providing valuable technological applications \cite{Gludovatz_hea_mech_properties}.}
{\bf III.} 
Maddox's scandal was only an apparent curse.
It was actually a blessing as it forced statistical analysis of the enthalpy versus entropy interplay.
Modeling can no longer ignore disorder --- often neglected due to compelling difficulties.
Advances in complex multi-component materials research will be driven by embracing disorder. It is unavoidable.

\section{Methods}

\noindent
Calculations of the enthalpy gain are performed using the \AFLOWCHULL\ module, using the command: \verb|aflow --chull --alloy=<system> --print=json|, where \verb|<system>| is a comma-separated list of the elements in the alloy system.
The enthalpy gain is contained in the \JSON\ output file under the keyword \verb|N+1_energy_gain|.
{To enhance statistical significance, only metallic alloy systems composed of the elements Ag, Al, Au, Ba, Be, Bi, Ca, Cd, Co, Cr, Cu, Fe, Ga, Hf, Hg, In, Ir, K, La, Li, Mg, Mn, Mo, Na, Nb, Ni, Os, Pb, Pd, Pt, Re, Rh, Ru, Sc, Sn, Sr, Ta, Tc, Ti, Tl, V, W, Y, Zn, Zr are considered as \protect\AFLOW\ data is much richer in that material domain.}
A total of {990 $N=2$, 14,190 $N=3$ and 148,675 $N=4$ phase diagrams are generated, with 202,261, 974,808 and 432,840}  {\it ab-initio} entries respectively.

\noindent
{\small  {\bf Acknowledgments.}
The authors thank Ohad Levy for fruitful discussions.
Research sponsored by DOD-ONR (N00014-15-1-2863, N00014-16-1-2326, N00014-17-1-2876).}

\noindent
{\small {\bf Competing interests:} The authors declare no competing
  interests. } 

\noindent
{\small  {\bf Author contributions.}
\noindent
  S.C. proposed the species-stabilization mechanism. 
  C.T. and C.O. developed the recursive stability codes.
  {D. H. generated the library of $N=4$ compounds.}
  All authors discussed the results and contributed to the writing of the article. 
} 

\noindent
{\small {\bf Data availability.}
All the \textit{ab-initio} data used in this analysis are freely available to the public as
part of the \AFLOW\ online repository and can be accessed through \AFLOW.org
following the \RESTAPI\ interface \cite{curtarolo:art92} and \AFLUX\ search language \cite{aflux}.
The \AFLOW\ source code is available for download at \verb|http://aflow.org/|
under GNU GPL version 3.}

\newcommand{\Ozolins}{Ozoli\c{n}\v{s}}

\end{document}